\documentclass[11pt,onecolumn,amssymb,nofootinbib]{revtex4}
\usepackage{amsmath, amsthm, amscd, amssymb}
\usepackage{bm,color}
\usepackage{bbm}

\begin{document}

\title{\bf The Continuous Spontaneous Localization Layering Effect from a Lattice Perspective}

\author{Stephen L. Adler}
\email{adler@ias.edu} \affiliation{Institute for Advanced Study,
Einstein Drive, Princeton, NJ 08540, USA.}

\author{Angelo Bassi}
\email{bassi@ts.infn.it}\affiliation{Department of Physics, University of Trieste, Strada Costiera 11, 34151 Trieste, Italy}
\affiliation{Istituto
Nazionale di Fisica Nucleare, Trieste Section, Via Valerio 2, 34127 Trieste,
Italy}

\author{Matteo Carlesso}
\email{matteo.carlesso@ts.infn.it}\affiliation{Department of Physics, University of Trieste, Strada Costiera 11, 34151 Trieste, Italy}
\affiliation{Istituto
Nazionale di Fisica Nucleare, Trieste Section, Via Valerio 2, 34127 Trieste,
Italy}

\begin{abstract}
For a solid lattice, we rederive the Continuous Spontaneous Localization (CSL) noise total energy gain of a test mass starting from a Lindblad formulation, and from a similar starting point rederive the geometry factor governing center of mass energy gain. We then suggest that the geometry factor can be used as a way to distinguish between low temperature cantilever motion saturation arising from CSL noise, and  saturation arising from thermal leakage.

\end{abstract}

\maketitle
\section{Introduction}

There has been much recent activity, both theoretical and experimental, involving the use of optomechanical systems to search for the noise postulated in Continuous Spontaneous Localization (CSL) models of state vector reduction~\cite{proposal-bah,propo-diosi,nimm,cant1,cant2,multi,goldwater,quantum-lim,rot,grav}. {Indeed, optomechanical systems can be exploited as extremely accurate noise detectors and therefore provide an important testbed for the CSL model. However, since no system is fully decoupled from its surrounding environment, also the latter acts on the system like a noise thus hindering a possible CSL signature. It is then important to devise a way to distinguish the two noisy contributions to better characterize the CSL one. }
 An important observation made by Nimmrichter, Hornberger, and Hammerer \cite{nimm} is that the center of mass diffusion rate for the cantilever involves a geometry dependent factor
$\tilde \mu(\vec k)$ that is defined as the Fourier transform of the classical mass density $\tilde\rho(\vec x)$.
Application of this geometry dependent factor to enhance the sensitivity of cantilever experiments by use of layered structures has been proposed \cite{multi} and applied \cite{multiexp}. {Such a geometrical dependence can be thus exploited to distinguish between possible CSL contributions and those from conventional environmental noises, as we will see below.}

In deriving this geometry dependent factor, the authors of \cite{nimm} assume that the cantilever is a homogeneous rigid body in which
excitation of internal degrees of freedom by the CSL noise can be neglected.  However, solids are actually lattices of molecules connected by intermolecular force ``springs'', in which internal excitations take
the form of phonon emission/absorption.  Calculations of heating of solids by CSL noise via phonon
excitation have been done by Adler and Vinante \cite{advin} and Bahrami \cite{bahr}, in both the white
and colored noise cases. These show that in the white noise case the heating rate depends only on the system mass $M$, and all dependence on the internal structure, which is present for colored or non-white noise, drops out.  In particular, there is no geometry dependent factor governing the total energy
excitation giving  the heating rate for white noise. One also obtains the same heating rate for a Fermi liquid \cite{fermions}.  In the colored noise case, a similar result is obtained, which additionally  depends on the noise spectrum, but there is no geometry dependence \cite{colored}. The aim of this  paper is to show how the
geometry dependent factor of \cite{nimm} arises when the phonon physics of realistic, non-rigid solids
is taken into account, by separating the lattice displacements used in \cite{advin,bahr} into center of mass and purely internal displacements.

The paper is organized as follows.  In Sec. 2 we repeat the calculation of the total heating rate done in \cite{advin}, by a different method that starts from the Lindblad equation for the density matrix. {Compared to the approach used in \cite{advin}, such a method is better suited for performing the calculations required in the next section.} In Sec. 3 we again start from the same Lindblad equation, and by separating the lattice displacements into center of mass and internal components, give a lattice physics derivation of the geometry
factor of \cite{nimm} in the white noise case.   Using this split, we  show that the total energy excitation splits
cleanly into a center of mass excitation, which is modulated by the absolute value squared  of the geometry factor, and an internal
energy excitation, which for large bodies accounts for most of the total energy excitation. In Sec. 3 we propose using the geometry
factor and the associated CSL layering effect as a way of experimentally distinguishing at low temperatures between genuine CSL noise effects,  and thermal leakages which can simulate CSL noise.

\section{Lindblad equation derivation of the total phonon heating rate}

In this section we calculate the total heating rate from phonon emission.  Instead of using the methods of \cite{advin,bahr,fermions}, we follow \cite{multi} and start from the CSL Lindblad type master equation, with the caret $\hat{}$ denoting operators,
\begin{equation}\label{lindblad}
\frac{d \hat \rho(t)}{dt}=-\frac{i}{\hbar}[\hat H,\hat \rho(t)] \, +  \, {\cal L}[\hat \rho(t)]~~~,
\end{equation}
where $\hat H$ is the Hamiltonian describing free evolution of the system, and
\begin{equation}\label{lind}
{\cal L}[\hat \rho(t)]=-\frac{\lambda}{2r_C^3\pi^{3/2} m_N^2} \int d^3z [\hat M(\vec z),[\hat M(\vec z),\hat \rho(t)]]~~~
\end{equation}
governs the CSL action on the system, with $m_N$ the nucleon mass, $\lambda$ the noise coupling, and $r_C$ the noise correlation length.
Assuming a mass-proportional CSL noise coupling, $\hat M$ is defined by
\begin{align}\label{mhat}
\hat M(\vec z)=&\int d^3x \,e^{-{(\vec z-\vec x)^2}/{2r_C^2}}\hat \rho(\vec x)~~~,\cr
\hat \rho(\vec x)=&\sum_\ell m_\ell \delta^3(\vec x-\hat x_\ell)~~~.
\end{align}
Here the sum over $\ell$ runs over all atoms of the system with mass $m_\ell$ and  position operator $\hat x_\ell$. By writing
\begin{equation}\label{fourier}
(2\pi)^{-3/2} r_C^{-3} e^{-\vec x^2/(2r_C^2)}= (2\pi)^{-3} \int d^3k \,e^{-r_C^2 \vec k^2/2 - i\vec k \cdot \vec x}~~~,
\end{equation}
and
\begin{equation}\label{hatmu}
\hat \mu(\vec k)=\int d^3 x \,e^{-i\vec k \cdot \vec x} \hat \rho(\vec x)
=\sum_\ell m_\ell e^{-i\vec k \cdot \hat x_\ell}~~~,
\end{equation}
a simple calcuation shows that ${\cal L}$ can be rewritten as
\begin{equation}\label{calLk}
{\cal L}[\hat \rho(t)]=-\frac{\lambda r_C^3}{2 \pi^{3/2} m_N^2} \int d^3 k \,e^{-r_C^2 \vec k^2}[\hat \mu(\vec k),[\hat \mu^\dagger (\vec k), \hat \rho(t)]]~~~.
\end{equation}

Let us now apply Eq. \eqref{calLk} to calculate the CSL energy gain rate $\Gamma$ :
\begin{equation}\label{gainrate}
\Gamma \equiv {\rm  Tr}\left( \hat H \frac {d\rho(t)}{dt}\right)= {\rm  Tr}\Big( \hat H {\cal L}[\hat \rho(t)]\Big)
=-\frac{\lambda r_C^3}{2 \pi^{3/2} m_N^2} \int d^3 k \,e^{-r_C^2 \vec k^2}{\rm Tr}\Big( \hat H  \,[\hat \mu(\vec k),[\hat \mu^\dagger (\vec k),\hat \rho(t)  ]] \Big) ~~~,
\end{equation}
where we have substituted Eq. \eqref{lindblad}.
Exploiting cyclic invariance of the trace,
the CSL energy gain takes the form
\begin{equation}\label{gainrate1}
\Gamma =-\frac{\lambda r_C^3}{2 \pi^{3/2} m_N^2} \int d^3 k \,e^{-r_C^2 \vec k^2} F(\vec k)~~~,
\end{equation}
where
\begin{equation}\label{defF}
F(\vec k)={\rm Tr}\Big( \hat \rho(t)  \,[\hat \mu^\dagger(\vec k),[\hat \mu (\vec k), \hat H ]] \Big).
\end{equation}

The next step is to evaluate the double commutator appearing in Eq. \eqref{defF} by introducing  phonon physics, following the exposition in the text of Callaway \cite{cal}.  We
consider the simplest case of a monatomic lattice with all $m_{\ell} $ equal to $m_A$, independent of the index $\ell$, and write the atom coordinate $\hat {\vec x}_{\ell}$ as
\begin{equation}\label{coord}
\hat {\vec x}_{\ell}=\vec R_{\ell}+ \hat{ \vec u}_{\ell}~~~,
\end{equation}
with $\vec R_{\ell}$ the equilibrium lattice coordinate and with $\hat{\vec u}_{\ell} $ the lattice displacement induced by the noise perturbation.   Writing
\begin{equation}\label{internsum}
\sum_{\ell} m_{\ell} e^{-i\vec k \cdot \hat {\vec x}_{\ell} }=m_A \sum_{\ell} e^{-i\vec k \cdot  \vec R_{\ell}}e^{-i\vec k \cdot  \hat{\vec u}_{\ell}  }~~~,
\end{equation}
we note that since the Gaussian in Eq. \eqref{gainrate} restricts the magnitude of $\vec k$ to be less than of order of $r_c^{-1}$, with $r_c \sim 10^{-5} {\rm cm}$, whereas
the magnitude of the lattice displacement is much smaller than $10^{-8}\,{\rm cm}$, the exponent in $e^{-i\vec k \cdot  \hat{\vec u}_{\ell}  }$ is a very small quantity.
So we can Taylor expand to write
\begin{equation}\label{expan}
e^{-i\vec k \cdot  \hat{\vec u}_{\ell}  }\simeq 1- i\vec k \cdot  \hat{\vec u}_{\ell} ~~~.
\end{equation}
Since this expression appears in a commutator, the leading term $1$ does not contribute to the energy gain rate.
We now substitute the expression \cite{cal} for the lattice displacement in terms of phonon creation and annihilation operators,
\begin{equation}\label{disp}
\hat{\vec u}_{\ell} = \frac{\Omega}{8\pi^3} \left(\frac{\hbar {\cal N}}{m_A}\right)^{1/2} \sum_j \int \frac{d^3 q}{(2\omega_j(\vec q))^{1/2}}
\big[\vec e^{\,(j)}(\vec q)   e^{i\vec q \cdot \vec R_{\ell}} \hat a_j(\vec q)+ \vec e^{\,(j)*}(\vec q) e^{-i\vec q \cdot \vec R_{\ell}} \hat a_j^{\dagger}(\vec q)\big]~~~,
\end{equation}
where the sum on $j$ runs over the acoustic phonon polarization states, and where $\Omega$ and ${\cal N}$ are respectively the lattice unit cell volume, and
 the number of unit cells.
The Hamiltonian $\hat H$ as well as $\hat \mu(\vec k)$  can  be expressed in terms of creation and annihilation operators of phonon modes \cite{cal},
\begin{equation}
\hat H=\frac{\mathcal N\Omega}{(2\pi)^3}\int d^3p\sum_i \hbar \omega_i(\vec p)\hat a_i^\dagger(\vec p)\hat a_i(\vec p).
\end{equation}
By imposing the commutation relations for the phonon operators $\left[\hat a_i(\vec q),\hat a^\dagger_j(\vec p)\right]=(2\pi)^3\delta_{ij}\delta^3(\vec q-\vec p) / (\mathcal N\Omega)$, one  finds that
\begin{equation}
F(\vec k)=-\frac{\hbar^2 m_A \Omega}{2(2\pi)^3}\sum_{l,l'} \int d^3p\sum_i\left|\vec k\cdot \vec e^{\,(i)}(\vec p)\right|^2\left(e^{-i(\vec k+\vec p)\cdot(\vec R_l-\vec R_{l'})}+e^{-i(\vec k-\vec p)\cdot(\vec R_l-\vec R_{l'})}\right).
\end{equation}
Exploiting the relation
\begin{equation}\label{sum1}
\sum_{\ell} e^{\pm i(\vec q-\vec k)\cdot \vec R_{\ell}}= \frac{8\pi^3}{\Omega}\delta^3(\vec q-\vec k)~~~,
\end{equation}
and taking into account the standard normalization of the Dirac delta:
\begin{equation}
\left[\delta^3(\vec q\pm\vec k)\right]^2=\delta^3(\vec q\pm\vec k)\int \frac{d^3x}{(2\pi)^3} e^{i(\vec q\pm\vec k)\cdot\vec x}=\delta^3(\vec q\pm\vec k) \frac{\mathcal N \Omega}{(2\pi)^3},
\end{equation}
$F(\vec k)$ reduces to
\begin{equation}\label{Fresult}
F(\vec k)=-\hbar^2 M \vec k^{\,2},
\end{equation}
where we employed the fact that $\vec k \cdot \vec e^{\,(j)}(\vec k)$ selects only the longitudinal acoustic phonon, thus giving $\sum_i\left|\vec k\cdot \vec e^{\,(i)}(\vec k)\right|^2=\vec k^{\,2}$, and we introduced the total mass $M=\mathcal Nm_A$.

By merging Eq. \eqref{Fresult}  with Eqs.~\eqref{gainrate}--\eqref{defF}, one finds the expression for the CSL energy gain,
\begin{equation}\label{stdform}
\Gamma=\frac{3}{4} \frac{\hbar^2 \lambda M}{m_N^2 r_c^2}~~~,
\end{equation}
where we used $\int d^3 k e^{-\vec k^2} \vec k^2 =\frac{3}{2} \pi^{3/2}$.
As emphasized in the Introduction, this formula depends only on the total mass $M$ and has no geometry dependent
factor. {Moreover, it can be shown that this result can be generalized to generic, not necessarily solid, systems \cite{fermions}. Conversely, the results below rely on the rigid body assumption. The CSL energy gain rate $\Gamma$ quantifies the heating of the CSL noise on the system. It can be exploited to determine experimental upper bounds on $\lambda$ as was done, e.g., in \cite{fermions} by comparing $\Gamma$ to the blackbody radiation emitted from a neutron star and Neptune.}

\section{Lattice derivation of the CSL geometry factor}

A salient feature of the calculation of the preceding section is that there has been no separation
of the center of mass displacement from the total site displacements $ \hat{ \vec u}_{\ell}$.  Phonon excitations, as
defined by Eq. \eqref{disp}, include both internal and center of mass displacements, and the energy
production rate of Eq. \eqref{stdform} is the sum of the center of mass  and the internal energy
production rates.

In this section
we isolate the center of mass excitation energy and the excitation of internal degrees of freedom. Our starting point is Eqs. \eqref{hatmu} and \eqref{calLk}, which express ${\cal L}$ in terms of the Fourier
transform of the mass density operator.  Substituting Eq. \eqref{coord} into Eq. \eqref{hatmu}, we get
\begin{equation}\label{subs}
 \hat \mu(\vec k)=\sum_\ell m_\ell e^{-i\vec k \cdot \vec R_\ell}
 e^{-i\vec k \cdot \hat{ \vec u}_{\ell}}~~~.
\end{equation}
Expanding the second exponential on the right as in Eq. \eqref{expan}, we get
\begin{equation}\label{hatmu1}
\hat \mu(\vec k)\simeq .....-i   \sum_\ell m_\ell e^{-i\vec k \cdot \vec R_\ell} \vec k \cdot \hat{ \vec u}_{\ell}~~~,
\end{equation}
where $....$ denotes c-number terms that do not contribute to the commutator in Eq. \eqref{calLk}.
We now use the transformation to center of mass and internal coordinates given in \cite{adram}, denoting
by $N$ the total number of atom sites $\ell$ (for a monatomic lattice, $N$ is equal to the number
of unit cells ${\cal N}$ introduced earlier)
\begin{align}\label{sep}
M=&\sum_{\ell=1}^N m_\ell~~~,\cr
\hat{\vec X}=&\sum_{\ell=1}^N m_\ell \hat{ \vec u}_{\ell}/M~~~,\cr
\hat{ \vec u}_{\ell}=&\hat{\vec  \xi}_\ell + \hat{\vec X}~~,~~~\ell=1,...,N-1~~~,\cr
\hat{ \vec u}_{N}=&\hat{\vec X}-\sum_{\ell=1}^{N-1} m_\ell \hat{ \vec u}_{\ell}/m_N~~~.
\end{align}
Under this transformation, the operator part of $\hat \mu(\vec k)$ splits into mutually commuting center of mass
and internal pieces,
\begin{align}\label{split}
\hat \mu(\vec k)=&\hat \mu(\vec k)_{\rm cm} + \hat \mu(\vec k)_{\rm int}~~~,\cr
\hat \mu(\vec k)_{\rm cm}=& -i\sum_{\ell=1}^N m_\ell e^{-i\vec k \cdot \vec R_\ell}\vec k \cdot \hat{\vec X}~~~,\cr
\hat \mu(\vec k)_{\rm int}=& -i\sum_{\ell=1}^{N-1} [ e^{-i\vec k \cdot \vec R_\ell}-
e^{-i\vec k \cdot \vec R_N} m_\ell/m_N ]\vec k \cdot \hat{\vec  \xi}_\ell~~~.
\end{align}
Defining the c-number geometry factor $\tilde \mu(\vec k)$ by
\begin{equation}\label{geom2}
\tilde \mu(\vec k)=\sum_{\ell=1}^N m_\ell e^{-i\vec k \cdot \vec R_\ell}~~~,
\end{equation}
we see that $\hat \mu(\vec k)_{\rm cm}$ is given by the geometry factor times $-i\vec k\cdot \hat{\vec X}$,
\begin{equation}\label{geom3}
\hat \mu(\vec k)_{\rm cm}=\tilde \mu(\vec k)(-i\vec k \cdot  \hat{\vec X})~~~.
\end{equation}

As shown in \cite{adram}, under the transformation of Eq. \eqref{sep}, the kinetic energy
part of $\hat H$ splits into a center of mass part and an internal part, and since the potential
energy part of $\hat H$  is a function of the internal coordinates only, we have
\begin{align}\label{hsplit}
\hat H=&\hat H_{\rm cm}+\hat H_{\rm int}~~~,\cr
\hat H_{\rm cm}=& -\hbar^2\frac{\hat {\vec \nabla}^2_X}{2M}~~~,\cr
\hat H_{\rm int}=&-\hbar^2\sum_{\ell=1}^{N-1}\frac{\hat {\vec \nabla}^2_{\xi_\ell}}{2m_\ell}+\frac{\hbar^2}{2M}
\left(\sum_{\ell=1}^{N-1}\hat {\vec \nabla}_{\xi_\ell}\right)^2+V(\hat{\vec  \xi}_1, ..., \hat{\vec  \xi}_{N-1})~~~.
\end{align}
Note that $\hat H_{\rm int}$ is not diagonal in the internal coordinates, which is why a center of mass
separation is not made when introducing phonons;  the phonon transformation of Eq. \eqref{disp} is
constructed to diagonalize the harmonic approximation to the lattice Hamiltonian.
Corresponding to the splitting of $\hat H$ in Eq. \eqref{hsplit}, the rates of increase of the center of mass and internal energy are given by
\begin{align}\label{gammasplit}
\Gamma_{\rm cm} = &{\rm Tr}\Big( \hat H_{\rm cm} {\cal L}[\hat \rho(t)]\Big)~~~,\cr
\Gamma_{\rm int}=&{\rm Tr} \Big(\hat H_{\rm int}  {\cal L}[\hat \rho(t)]\Big)~~~.
\end{align}
Combining this with Eqs.~\eqref{calLk} and \eqref{split}, and using the cylic property of the trace
to throw the commutators onto the Hamiltonian factor, we get
\begin{align}\label{throw}
\Gamma_{\rm cm} = &-\frac{\lambda r_C^3}{2 \pi^{3/2} m_N^2} \int d^3 k \,e^{-r_C^2 \vec k^2}|\tilde \mu(\vec k)|^2
{\rm Tr}\Big([\vec k \cdot \hat{\vec X},[\vec k \cdot \hat{\vec X},\hat H_{\rm cm} ]]  \hat \rho(t)\Big)~~~,\cr
\Gamma_{\rm int} = &-\frac{\lambda r_C^3}{2 \pi^{3/2} m_N^2} \int d^3 k \,e^{-r_C^2 \vec k^2}
{\rm Tr}\Big([\hat \mu^\dagger(\vec k)_{\rm int},[\hat \mu(\vec k)_{\rm int},\hat H_{\rm int} ]]  \hat \rho(t)\Big)~~~.
\end{align}
The double commutator in $\Gamma_{\rm cm}$ is easily evaluated to give the c-number
\begin{equation}\label{doublecomm}
[\vec k \cdot \hat{\vec X},[\vec k \cdot \hat{\vec X},\hat H_{\rm cm} ]]=- \hbar^2\vec k^2/M~~~,
\end{equation}
and using  ${\rm Tr} \hat \rho(t)=1$ we
get for the center of mass energy excitation
\begin{equation}\label{cmen}
\Gamma_{\rm cm} = \frac{\lambda r_C^3\hbar^2}{2M \pi^{3/2} m_N^2} \int d^3 k \,e^{-r_C^2 \vec k^2}  \vec k^2 |\tilde \mu(\vec k)|^2~~~.
\end{equation}
{Similarly to the total energy gain $\Gamma$, also $\Gamma_{\rm cm}$ can be directly exploited to determine upper bounds on $\lambda$. Indeed, $\Gamma_{\rm cm}$ determines the rate at which the CSL noise excites the center of mass motion, which can be suitably measured through, e.g., optomechanical tests \cite{cant1,grav,cant2,multiexp,roomtemp,opto}.}
If the geometry factor $\tilde \mu(\vec k)$ were equal to $M$, this would reduce to Eq. \eqref{stdform},
but Eq. \eqref{geom2} implies that in general  $|\tilde \mu(\vec k)|\leq M$, so the center of mass
energy excitation is always smaller than the total energy excitation.   We do not attempt to evaluate
$\Gamma_{\rm int}$ from Eq. \eqref{throw}; the simplest way to calculate it  is from the difference
$\Gamma-\Gamma_{\rm cm}$, both terms of which are given by relatively simple formulas.

Defining the classical mass density by
\begin{equation}\label{class}
\tilde \rho(\vec x)=\sum_{\ell=1}^N   m_\ell \delta^3(\vec x-\vec R_\ell)~~~,
\end{equation}
the geometry factor can be written as
\begin{equation}\label{class1}
\tilde \mu(\vec k)=\int d^3 x  \,e^{-i\vec k \cdot \vec x} \tilde \rho(\vec x) ~~~,
\end{equation}
which is the Fourier transform of the classical mass density.  The formula of Eq. \eqref{class1} is
used in \cite{nimm} and \cite{multi} to calculate $\tilde \mu(\vec k)$ for various cantilever geometries.

\section{Use of the geometry factor to distinguish CSL noise from thermal leakage}

We conclude by noting that the geometry factor dependence of the center of mass excitation energy
 gives a way of distinguishing between low temperature thermal saturation resulting from CSL
noise, and thermal saturation resulting from thermal leakage.  It seems plausible that thermal saturation
resulting from thermal leakage will only depend on the mass ratios of different materials in the cantilever
test mass, and not on the precise test mass geometry. {Indeed, in a system such as \cite{multiexp}, the thermal noise leads to thermal energy gain $\Gamma_\text{\tiny th}=\gamma_\text{\tiny th} k_\text{\tiny B} T$ \cite{multi}, where $\gamma_\text{\tiny th}$ is the damping rate and $k_\text{\tiny B}$ and $T$ are respectively the Boltzmann constant and the environmental temperature.
In particular, $\gamma_\text{\tiny th}$ can strongly depend on the experimental setup and on the materials  of which the system is made, while it does not depend on its geometry.} Moreover, this is an assumption that can be
tested in auxiliary experiments in which a large thermal leakage is introduced to the cantilever.  On the other
hand, as shown in \cite{multi}, the CSL noise sensitivity of the cantilever is strongly dependent on the
geometry of the test mass;   by constructing the test mass from alternating layers of different materials
the CSL sensitivity can be enhanced. {As was shown in Figure 6 of \cite{multi}, by fixing a value of the mass of the system and employing a different number of layers (consequently their thickness also changes since the total mass is fixed), one can fine-tune the sensitivity to CSL for different values of $r_C$ corresponding to roughly the thickness of the layer.} Thus by performing a cantilever experiment with several test
masses with identical mass ratios of materials (and hence, under the assumption made above, with identical
sensitivities to thermal leakage), but different layering geometries with significantly different CSL noise sensitivities, it should be possible to distinguish a true CSL signal from a thermal leakage background.

\section{Acknowledgements}

S.L.A. acknowledges  the hospitality of the Aspen Center for Physics,
which is supported by the National Science Foundation under grant PHY-1607611.
AB acknowledges financial support from FQXi, the COST Action QTSpace (CA15220), INFN, and hospitality from the IAS Princeton, where part of this work was carried out.
AB and MC acknowledge financial support from the H2020 FET Project TEQ (grant n.~766900).

\end{document}